\documentstyle[aps,prb,multicol,epsfig]{revtex}
\begin{document}
\title{ Finite size and temperature effects in  the $J_1-J_2$ model on a strip}
\author{ E.S. Pisanova}
\address{Faculty of Physics, The University of Plovdiv,
Tzar Assen 24, 4000 Plovdiv, Bulgaria}

\author{N.B. Ivanov  and N.S. Tonchev\cite{tonchev}}
\address{Institute  of Solid  State
Physics, Bulgarian Academy of Sciences,
Tzarigradsko
chaussee 72, 1784 Sofia, Bulgaria}
\date{\today}
\maketitle
\begin{abstract}
Within Takahashi's spin-wave theory  we study finite size and temperature effects
near the quantum critical point in  the    $J_{1}-J_{2}$
Heisenberg antiferromagnet defined on a strip ($L\times\infty$).
In the continuum limit, the theory predicts universal finite  size and temperature corrections
and describes  the  dimensional crossover  in  magnetic properties
from 2+1  to 1+1 space-time dimensions.
\end{abstract}
\draft
\begin{multicols}{2}
Takahashi's spin-wave theory is a useful tool for  qualitative study of   thermodynamic
properties in low-dimensional  spin  systems ~\cite{takahashi1}.  For instance,  the
theory   is capable to reproduce   the universal
low-temperature behavior in two-dimensional (2D)   Heisenberg
antiferromagnets (HAFM)~ \cite{takahashi2}  originally predicted on the basis of
the 2+1 nonlinear $\sigma$ model~\cite{chakravarty} in the limit
$\rho_s\ll  T$ ($\rho_s$  is  the spin-wave stiffness constant). Near the quantum
critical point  2D HAFM  is expected to possess  universal magnetic properties
for arbitrary $T/\rho_s$ provided $\rho_s,T\ll J$,  $J$ being  a
characteristic exchange energy ~\cite{chubukov}.
Finite size and temperature effects in  2D HAFM  in the limit
$\rho_s\ll  T$  have  previously  been analyzed in the geometry
$L\times L$~\cite{hasenfratz1}.

  In the present work we apply Takahashi's
approach to a  frustrated $J_1-J_2$
Heisenberg antiferromagnet defined on a  strip with linear
dimensions  $L\times\infty$ . In this   geometry
the coupled self-consistent  equations
for the spectral gap $\Delta$,  the  current  field   $g=-\langle
a^{\dag}_{\bf 0} b^{\dag}_{\hat{x}}\rangle$,  and the density  field  $f=\langle a^{\dag}_{\bf 0}
a_{\hat{x}+\hat{y}}\rangle$  can be represented in the form~\cite{ivanov}
\begin{eqnarray}\label{gf}
S+\frac{1}{2}&=&\frac{1}
{2L}\sum_{k_1}   \int_{-\pi}^{\pi}\frac{dk_2}{2\pi}
\frac{1}
{\sqrt{1-\eta_{\bf k}^{2}\gamma_{\bf k}^{2}}}
\coth{\left( \frac{\epsilon_{\bf k}}{2T}\right) },\nonumber \\
g=S+\frac{1}{2}&-&\frac{1}
{2L}\sum_{k_1}   \int_{-\pi}^{\pi}\frac{dk_2}{2\pi}
\frac{1-\gamma_{\bf k}^{2}\eta_{\bf k}}
{\sqrt{1-\eta_{\bf k}^{2}\gamma_{\bf k}^{2}}}
\coth{\left(\frac{\epsilon_{\bf k}}{2T}\right)},\nonumber \\
f= S+\frac{1}{2}&-&\frac{1}
{2L}\sum_{k_1}   \int_{-\pi}^{\pi}\frac{dk_2}{2\pi}
\frac{1-\kappa_{\bf k}}
{\sqrt{1-\eta_{\bf k}^{2}\gamma_{\bf k}^{2}}}
\coth{\left(\frac{\epsilon_{\bf k}}{2T}\right)},
\end{eqnarray}
where  $\epsilon_{\bf k}=4J_{1}g\eta_{\bf k}^{-1}\sqrt{1-\eta_{\bf k}^{2}\gamma_{k}^{2}}$
is the magnon dispersion relation [$\eta_{\bf k}^{-1}=1-(\alpha f/g)
(1-\kappa_{\bf k})+\mu$,  $\kappa_{\bf k}=\cos k_1\cos k_2$,
$\gamma_{\bf k}=(\cos k_1+\cos k_2)/2$] $\alpha
=J_2/J_1$ is the frustration parameter, and $S$ is the value of the lattice
spin. The components of the wave vector ${\bf k}=k_1{\bf
\hat{x}}+ k_2 {\bf \hat{y}}$  take the values  $k_{1}=2\pi n_{1}/L$
[$n_{1}=0,\pm 1, \ldots ,\pm (N_{0}-1)/2$] and
$k_{2} \in [-\pi,\pi]$, where $L$ is an odd integer.
 We use a system of  units where
$\hbar = k_B=a=1$, $a$ is the lattice spacing.
The chemical potential $\mu$ is connected  to the spectral gap by the
relation  $\Delta = 4J_1g\sqrt{2\mu}$\, .

At $T=0$  and in an infinite geometry ($L=\infty$) the system  is
in the  magnetic N\'{e}el  state, characterized by  the on-site magnetization
\begin{equation}
\label{mgn}
m_0=S+\frac{1}{2}-\frac{1} {2}\int_{-\pi}^{\pi}\int_{-\pi}^{\pi}
\frac{dk_{1}dk_{2}}{(2\pi)^2}
\frac{1}{\sqrt{1-\eta_{\bf k}^{2}\gamma_{\bf k}^{2}}},
\end{equation}
up  to the quantum critical point $\alpha_c$.  Recent studies
 predict that the sysrem is
 magnetically disordered from  $\alpha_c=0.38$ up to  $\alpha \approx 0.6$,
 and support the hypothesis that $\alpha_{c}$ is a second-order
 phase transition point described by the $O(3)$ nonlinear $\sigma$ model
 ~\cite{shushkov}.
Eqs.~(\ref{gf}) are supposed to describe the magnetically disordered phase
at finite temperatures both in the infinite $\infty\times\infty$ and
in the strip  $L\times\infty$  geometries . In addition, the theory can
also be applied  for the  magnetically disordered ground state of the strip
system.  In both cases  the excitation gap $\Delta  \neq 0$ appears as a
result of the constraint that implies  zero total staggered magnetization
[the first of
Eqs.~(\ref{gf})] and the theory is restricted to strip systems
with the integer "rung" spin $SL$.  An equivalent spin-wave description  of
the ground state of finite-size systems has originally been proposed
by Hirsch and Tang. ~\cite{hirsch}

Leading corrections
in $T$ and $1/L$ near the quantum critical point $(T,\alpha)=(0,\alpha_c)$
can be obtained in  the low-energy long-wavelength limit of the theory
when  the magnon dispersion relation takes the form
\begin{equation}\label{epsilon}
\epsilon_{\bf k}=c\sqrt{M^2+k^2}.
\end{equation}
Here $c=4J_1g\beta $ is the spin-wave velocity $\left(\beta =
\sqrt{1/2-\alpha f/g}\right) $  and $M=\sqrt{2\mu}/\beta$ is the magnon "mass"  which is connected to the
spectral gap by the relation $\Delta=cM$. In the framework of the discussed theory the
spin-wave stiffness constant is defined by the relation $\rho_s=2J_1g\beta^2m_0$.
In the long-wavelength limit Eqs. (1) are simplified  to
\begin{eqnarray}\label{tak2}
S+\frac{1}{2}-f+g&=&\frac{1}{\beta}W(L,T),\nonumber \\
S+\frac{1}{2}-g&=&\frac{2\beta^{2}+1}{4\beta}U(L,T)+\frac{\beta}{2}M^2W(L,T),\nonumber \\
S+\frac{1}{2}-f&=&\frac{1}{2\beta}U(L,T),
\end{eqnarray}
where
\begin{eqnarray}\label{wu}
W(L,T)&=&\frac{2T}{cL}\sum_{n=-\infty}^{\infty}
\sum_{k_{1}}\int_{-\pi}^{\pi}
\frac{dk_2}{2\pi}\frac{1}{ \omega_n^2/c^2+M^{2}+k^{2}}, \nonumber \\
 U(L,T)&=&\frac{2T}{cL}\sum_{n=-\infty}^{\infty}
\sum_{k_{1}}\int_{-\pi}^{\pi}
\frac{dk_2}{2\pi}\frac{k^2}{ \omega_n^2/c^2+M^{2}+k^{2}}.
\end{eqnarray}
In the last equations $\omega_n=2\pi nT$ and $k=\sqrt{k_1^2+k_2^2}$.\\

{\em  Finite-size effects at T=0:}\\

In the ground state  the functions  $W_0(L)\equiv W(L,0)$  and
$U_0(L) \equiv U(L,0)$ take the forms
$$
W_0(L)=\frac{1}{L}\sum_{k_{1}}\int_{-\pi}^{\pi}\frac{dk_2}{2\pi}
\frac{1}{\sqrt{M^{2}+ k^2}}
$$
and
$$
U_0 (L)=\frac{1}{L}\sum_{k_{1}}\int_{-\pi}^{\pi}\frac{dk_2}{2\pi}
\sqrt{M^{2}+ k^2} -M^{2}W_{0}(L).
$$
To extract the  finite-size corrections  we follow  the method
suggested in Ref.~\onlinecite{brankov} (see also Ref~\onlinecite{BDT}).
  Using integral representations for
the integrands\cite{chamati},  the above  two equations
 are, respectively,  transformed to
$$
W_0(L)=\int_{0}^{\infty}\frac{dt}{2\pi} \frac{e^{-M^2t}}{t}
\mathrm{erf}(\pi \sqrt{t})Q_{L}(t)
$$
and
\begin{eqnarray}
U_0(L)&=&M+\int_{0}^{\infty}\frac{dt}{2\sqrt{\pi}} \frac{e^{-M^2t}}{t^{3/2}}
\left[1- \frac{\mathrm{erf}(\pi \sqrt{t})}{\sqrt{4\pi t}}Q_{L}(t)\right]\nonumber\\ 
&-&M^{2}W_{0}(L),\nonumber
\end{eqnarray}
where $\mathrm{erf}(z)$ is the error function and
$Q_{L}(t)=(1/L)\sum_{n=-L/2+1/2}^{L/2-1/2}
\exp [-(2\pi n/L)^2t]$.  Note that $\lim_{L\to\infty}{Q_{L}(t)}=
\mathrm{erf}(\pi\sqrt{t})/\sqrt{4\pi t}$.

The finite-size contributions in
$W_0(L)=W_0(\infty)+\delta W_0(L)$ and  $U_0(L)=U_0(\infty)+\delta U_0(L)$
can  be separated by using   the  asymptotic formula $(L>>1)$
(see Ref.~\onlinecite{BDT}, p. 148)\cite{new}
\begin{equation}
Q_{L}(t)\cong \frac{1}{\sqrt{4\pi t}}\left[\mathrm{erf}(\pi t^{1/2})
+ 2\sum_{l=1}^{\infty}
\exp(-l^{2}L^{2}/4t)\right].
\label{asimp}
\end{equation}
 The result reads
\begin{eqnarray}
\delta W_0(L)&=&\frac{M}{\sqrt{\pi}} \sum_{l=1}^{\infty} \int_{0}^{\infty}
\frac{dt}{2\pi}t^{-3/2}\nonumber \\
&\times& \exp
\left(-t-\frac{l^{2}L^{2}M^{2}}{4t}\right)\mathrm{erf}\left(\frac{\pi\sqrt{t}}
{M}\right),\nonumber
\end{eqnarray}
\begin{eqnarray}
\delta U_0(L)&=&-\frac{M^{3}}{\sqrt{\pi}}
\sum_{l=1}^{\infty}\int_{0}^{\infty}
\frac{dt}{4\pi}t^{-5/2}\nonumber \\
&\times& \exp{\left(-t-\frac{l^{2}L^{2}M^{2}}{4t}\right)}
\mathrm{erf}\left(\frac{\pi\sqrt{t}}{M}\right)
- M^{2}\delta W_0(L).
\end{eqnarray}

For $ M<<1$  the error function in the above equations can  be
replaced by unity  and after some algebra one gets
\begin{equation}
\delta W_0(L)=\frac{1}{\pi L}\mathrm{Li}_{1}\left( e^{-LM}\right)
\end{equation}
and
\begin{equation}
\delta U_0(L)=-\frac{1}{\pi L^3}F(LM).
\end{equation}
Here $\mathrm{Li}_{s}(z)=\sum_{n=1}^{\infty}z^{n}/n^{s}$
 is the polylogarithmic function of $s$ kind and
\begin{equation}
F(y)\equiv y^{2}\mathrm{Li}_{1}(e^{-y}) +y\mathrm{Li}_{2}(e^{-y})
+\mathrm{Li}_{3}(e^{-y}).
\end{equation}

Using  Eq.~(\ref{mgn}),     near the quantum critical point $(T,\alpha)=(0,\alpha_{c})$
the first of equations (\ref{tak2}) takes the form
\begin{equation}\label{gap}
cM=\frac{2c}{L} {\rm arcsinh}
\left[ \frac{1}{2}\exp \left(-\frac{2\pi\rho_sL}{c}\right)\right].
\end{equation}

Finally, a subtraction of the  $L=\infty$ contributions in the last two of
 Eqs. (\ref{tak2})
leads to the expressions
\begin{eqnarray}\label{gapg}
g&=&g_c+\frac{(2\beta_c^{2}+1)}{4\pi\beta_c}
\frac{1}{L^{3}}F(LM)-  \frac{(2\beta_c^{2}-1)}{4\beta_c^2}
U_0(\infty)(\beta-\beta_c)\nonumber \\
&-&\frac{\beta_c}{2}M^2\delta W_0(L)
\end{eqnarray}
and
\begin{equation}\label{gapf}
f=f_c+\frac{1}{2\pi\beta_c}\frac{1}{L^{3}}F(LM)+
\frac{U_0(\infty)}{2\beta_c^2}(\beta-\beta_c).
\end{equation}
Here $g_c$, $f_c$, and $\beta_c$ are the values of the respective functions at
the quantum critical point.  The above expressions are valid near the quantum critical point.

Eq.~(\ref{gap}) exactly reproduces the saddle point
equation  (with an appropriate regularization)
in the $1/N$ expansion of the $O(N)$ nonlinear sigma model in
2+1 space-time dimensions. ~\cite{tsvelik}
As can be expected from the relativistic dispersion
relation (\ref{epsilon}), the ratio $c/L$ plays the role of an
effective temperature.  It is well known that the gap  equation  (\ref{gap})
describes three different regimes, i.e.,  (i) the renormalized classical,
(ii) the quantum critical,  and (iii) the quantum disordered  regimes
(see, e.g.,  Ref~\onlinecite{BDT}). \\
(i) For $\rho_sL/c\gg1$ (large linear-size behavior)
the approximate solution  reads $\Delta\equiv
Mc=(c/L)\exp (-2\pi\rho_sL/c)$.
Using the mentioned symmetry and the refined result~\cite{hasenfratz2}
 for the correlation length $\xi$ at low temperatures, the expression
 for $\Delta$ can be improved to
\begin{equation}\label{dl}
\frac{\Delta(L)}{c} =A\exp\left(-\frac{2\pi\rho_sL}{c}\right)\left(
1-\frac{c}{4\pi\rho_ sL}\right),
\end{equation}
where $A$ is some dimensionless parameter.
The last equation may be interpreted as the spectral gap
of a 1D HAFM defined on a  ladder
characterized by  the  integral   "rung" spin
$S_{eff}=\rho_sL/c$. \cite{rezende}\\
(ii)   Exactly at the quantum critical point  Eq. (\ref{gap}) gives
\begin{equation}
\Delta=y_0\frac{c}{L}\, ,
\end{equation}
where $y_0=2\ln (1/2+\sqrt{5}/2)$ is an universal constant.\\
 Eqs. (\ref{gapg}) and (\ref{gapf}) lead to  the   universal finite-size contribution for
the internal energy per spin [$e(L)=-2J_1(g^2-\alpha f^2)$]
\begin{equation}
e(L)=e_c(\infty )-\frac{c}{\pi L^3}\tilde{F}(LM),
\end{equation}
where $\tilde{F}(y)=F(y)-(y^2/2){\mathrm Li}_1({\mathrm e}^{-y})$.
It can be shown that for $\alpha < \alpha_c$ and in the limit $LM \ll 1$ the
function $\tilde{F}(LM)$ reduces to $\zeta (3)$,  whereas exactly at the critical point
we have $\tilde{F}=4\zeta (3)/5 +y_{0}^{3}/12$: the last result is obtained
by using  Sachdev's  identity  (Ref.~\onlinecite{sachdev}).
Here $\zeta (x)$ is the Riemann zeta function.

(iii)Finally,  in the quantum disordered regime ($\alpha >\alpha_c$) we get
\begin{equation}
e(L)=e_c(\infty)-\frac{c M^2}{2\pi L}\exp(-LM).
\end{equation}
The exponential form of the finite-size corrections reflects the existence of an
excitation gap in the quantum disordered phase.
Notice, however,  that in this regime
the system is known to possess gapped triplet
excitations ~\cite{chakravarty},
whereas Takahashi's theory is entirely based  on transverse
spin-wave excitations.

{\em Finite-size effects at low temperatures:}\\

Using the Jacobi identity
$$\sum_{m=-\infty}^{\infty}\exp{(-um^{2})}=
\sqrt{\frac{\pi}{u}}\sum_{m=-\infty}^{\infty}
\exp{\left(-\frac{\pi^{2}m^{2}}{u}\right)}
$$
and the Poisson summation formula, one can transform
Eqs.~(\ref{wu}) into the forms $W(L,T)=W(\infty,0)+\delta
W(L,T)$ and $U(L,T)=U
(\infty,0)+\delta  U(L,T)$. In what follows we
will consider only finite size and temperature
corrections to the internal energy. Since at low temperatures
the magnon mass $M$ is exponentially small,
it is enough to calculate the function  $\delta  U(L,T)$  for $M=0$.
The result reads
\begin{equation}\label{u}
\delta U(L,T)_{M=0}=\frac{2\Gamma(3/2)}{\pi^{3/2}c^3}T^3
\sum_{m,n=-\infty}^{\infty} \left( m^{2}+n^{2}L^{2}T^{2}/c^2\right)^{-3/2},
\end{equation}
 where the term $m=n=0$ must be omitted.\\
(i) In the regime $T\gg c/L$  it is convenient to transform Eq.~(\ref{u})
to the form
\begin{eqnarray}\label{id}
\delta U(L,T)_{M=0}&=&\frac{2\zeta(3)T^{3}}{\pi c^{3}}+ \frac{4\zeta(2)T}
{\pi c L^{2}}\nonumber \\
&+&\frac{16T^{2}}{c^{2}L}\sum_{l,m=1}^{\infty}\frac{l}{m}
K_{1}\left(\frac{2\pi lmLT}{c}\right),
\end{eqnarray}
[$K_{1}(x)$ is the McDonald function] which
yields the following expression for the internal energy
per spin:
\begin{eqnarray}
\label{T3}
e(L,T)&=&e_c(\infty,0)+ \frac{2 \zeta (3)}{\pi c^2}\left\{ 1 +
O\left[(TL/c)^{-3/2}e^{-2\pi \frac{TL}{c}}\right]\right\}\nonumber \\
&\times& T^3+
\frac{4\zeta (2)}{\pi L^2}T.
\end{eqnarray}
In the considered long-wavelength approximation the expansion
parameter $(TL/c)^{-1}$
gives exponentially small contributions in the $T^{3}$ prefactor. In
the bulk limit the $T^{3}$ term in Eq.~(\ref{T3})  exactly reproduces
the  expected contribution   to the internal energy  of the 2D HAFM
from the Goldstone modes at ${\bf k}=(0,0)$ and ${\bf k}=(\pi ,\pi )$
(see, e.g.,  Ref.~\cite{hasenfratz1}).

(ii) In the regime $T\ll c/L$  it is convenient
to use instead of Eq. ~(\ref{id}) the expression
\begin{eqnarray}
\delta U(L,T)_{M=0}&=&\frac{2\zeta(3)}{\pi L^{3}}+ \frac{4\zeta(2)T^{2}}
{\pi c^{2} L}\nonumber\\
&+&\frac{16T}{cL^{2}}\sum_{l,m=1}^{\infty}\frac{l}{m}
K_{1}\left(\frac{2\pi lmc}{LT}\right)
\end{eqnarray}
which leads to  the   formula
\begin{eqnarray}
e(L,T)&=&e_c(\infty,0)+\frac{2c\zeta (3)}{\pi }\left\{1 +
O\left[(TL/c)^{3/2}e^{-2\pi \frac{c}{TL}}\right]\right\}\nonumber\\
&\times&\frac{1}{L^3}+
\frac{4\zeta (2)}{\pi cL}T^2.
\end{eqnarray}

In conclusion,  based on Takahashi's spin-wave theory
we have obtained expressions for the spectral gap and
the internal energy per spin of a  frustrated Heisenberg antiferromagnet
defined on a strip. The expressions are valid in the continuum limit close to 
the quantum critical
point and have a universal form depending only on the macroscopic
magnetic parameters $\rho_s$ and $c$.

\acknowledgments
This work was supported by the Bulgarian Science Foundation
(Grant No. 817/98)  and the University of Plovdiv (Contract 01-F-23).
N.B.I. and N.S.T. acknowledge the financial support, respectively,
from  the Max-Planck Institut f\"ur Physik Komplexer Systeme (Dresden, Germany)
and the  Associate and Federation
Scheme  of ICTP (Trieste, Italy).
 
\end{multicols}{2}
\end{document}